\documentclass[twocolumn,showpacs,preprintnumbers,amsmath,amssymb]{revtex4}
\usepackage{graphicx}% Include figure files
\usepackage{dcolumn}% Align table columns on decimal point
\usepackage{bm}% bold math

\def\bra#1{ \langle #1\,\mid}
\def\ket#1{ \mid\!#1\rangle}
\def\inn#1#2{\langle #1\mid #2\rangle}

\begin{document}

\title{Concurrence Vectors for Entanglement of High-dimensional Systems}

\author{You-Quan Li and Guo-Qiang Zhu}
\affiliation{Zhejiang Institute of Modern Physics, Zhejiang
University, Hangzhou 310027, P. R. China}

\begin{abstract}
The concurrence vectors are proposed by employing the
fundamental representation of $A_n$ Lie algebra, which
provides a clear criterion to evaluate the
entanglement of bipartite system of arbitrary
dimension for both pure and mixed states.
Accordingly, a state is separable if the
norm of its concurrence vector vanishes. The state
vectors related to SU(3) states and SO(3) states are
discussed in detail. The sign situation of nonzero
components of concurrence vectors of entangled bases
presents a simple criterion to judge whether the whole
Hilbert subspace spanned by those bases is entangled,
or there exists entanglement edge. This is illustrated in
terms of the concurrence surfaces of several concrete examples.
\end{abstract}

%03.65.Ud Entanglement and quantum nonlocality
%03.67.Mn Entanglement production, characterization and manipulation
%03.65.Ca Formalism

\received{15 September 2003, revised 31 July 2004}%

\maketitle

\section{Introduction}

Entanglement, as one of the most intriguing features of quantum
system, has been a subject of much study in recent years.
It is regarded as a valuable resource in quantum computation and
communication processions, which allows quantum
physics to perform tasks that are classically impossible
\cite{someReviews}. The qubit
%, a quantum system with a two-dimensional Hilbert space,
used to be considered as the building blocks of quantum computers,
and the entanglement of bipartite systems of qubits has been well studied.
Most recently, moreover, an arbitrary polarization state of a
single-mode biphoton was considered \cite{Zhukov} to generate
qutrit which is a system whose states constitute a
three-dimensional Hilbert space. Experimentally, a new technique
has been introduced to generate and control entangled
qutrits\cite{Thew}, which reveals a source capable of generating
maximally entangled states with a net state fidelity. The qutrit
system was also suggested to be realized by nuclear magnetic
resonance (NMR) utilizing deuterium nuclei partially oriented in
liquid crystalline phase \cite{Kumar} or by trapped ions
\cite{Klimov}. The use of qutrits instead of qubits was shown
\cite{Bruss} to be more secure against symmetric attacks on a
quantum key distribution protocol while the violation of local
realism for two maximally entangled $N$-dimensional state is
stronger than for qubits\cite{Kaszlikowski}. Thus an effective measurement
of entanglement for high dimensional Hilbert space become useful.

As is known that Peres \cite{Peres96} proposed the
partial transposition criterion as a necessary
condition for separability (independent of the
dimension of the state), which was later shown to be
sufficient for the 2 by 2 and 2 by 3 cases by
Horodecki {\it et. al.} \cite{MHorodecki96}. More
authors carried out further discussions and proposed
various procedures\cite{Rungta01,Badziag02}
for the similar purpose. Since the concurrence
introduced early by Hill and Wootters
\cite{HillWootters} is an important evaluation of
entanglement for qubits, an systematical extension for
qutrits as well as to high-dimensional system
should be helpful.

In this paper we present an extension of Hill-Wootters'
concurrence on the basis of roots of $A_n$ Lie algebra.
In next section, we define a concurrence vector whose
norm can be used to evaluate the entanglement
of pure states, i.e., a separable state has a
vanishing norm. Its relation to other
entanglement measurement is also discussed.
In section \ref{sec:mix}, we extend the concurrence vector
to mixed states.
In section
\ref{sec:surface}, we discuss the concurrence
surface for several concrete cases, which is
expected to be helpful for the study of entanglement
evolution. The advantage of our proposal is that one
can easily judge whether a Hilbert subspace spanned by
some entangled states is fully entangled, or there
exists entanglement edge in the subspace.
In the last section, a brief summary and acknowledgement are given.  %

\section{Concurrence vector for pure states: from qubits to qudits}\label{sec:vetor}

\subsection{Formal extension}

Let us recall the ``concurrence'' introduced by Hill and Wootters
\begin{eqnarray}
&&C(\psi)=|\langle\psi\mid\tilde{\psi}\rangle|
  \nonumber\\
&&\mid\tilde{\psi}\rangle=(\sigma_y\otimes\sigma_y)\mid\psi^*\rangle,
\label{eq:concurrence}
\end{eqnarray}
which provides an easy evaluation of entanglement for a pair of
qubits. The qubit is a category of two-dimensional Hilbert space
well described by Pauli matrices that carry out the fundamental
representation of group SU(2). It is convenient to choose the
SU(2) generators as $J_z$, $J_+$ and $J_-$ satisfying
\begin{equation}
[J_z,\,J_\pm ]=\pm J_\pm, \;\; [J_+,\, J_- ]=2 J_z.
 \label{eq:su2}
\end{equation}
In terms of these operators, Hill-Wootters' definition
(\ref{eq:concurrence}) can be equivalently replaced by
$\mid\tilde{\psi}\rangle=(J_+ - J_-)\otimes(J_+ -
J_-)\mid\psi^*\rangle$. This expression can be conveniently
extended to the case of high-dimensional Hilbert space because
$\{J_x,\,J_y\,J_z\}$ do not has counterpart for high rank groups
but $\{J_+,\,J_-\,J_z\}$ has.
We need to describe a qutrit by the generators of
SU(3) group. Furthermore, the SU(N) group is required
for the case of $N$-dimensional Hilbert
space. Hereafter, we will adopt the standard
terminology in group theory in order to avoid possible
ambiguities, meanwhile keeping physics picture  as
much as possible. Some helpful mathematical concepts
are given in the appendix.

Let us analyzes the separability of the bipartite state $\ket{\psi}$.
Supposing it is separable, we will have
$\ket{\psi}=\ket{\phi_A}\otimes\ket{\phi_B}$
and generally
$\ket{\phi_A}=\sum_{m=1}^N
a_m \ket{m}$;
$\ket{\phi_B}=\sum_{m=1}^N
b_m \ket{m}$. The action of $E_\alpha$ on the state
$\ket{\phi_A}$ (similar to $\ket{\phi_B}$ )
makes one term in the summation not vanish
merely. This is because $E_\alpha$ maps some one
state saying $\ket{m'}$ to another one saying
$\ket{m''}$ but the others to null. As a result, we have
$\langle\phi_1\mid E_\alpha\ket{\phi^*_A}=
a^*_{m''}a^*_{m'}$.
Employing the operator related to the corresponding negative
root, we have $\langle\phi_1\mid E_{-\alpha}\ket{\phi^*_A}=
a^*_{m'}a^*_{m''}$.
Referring to Hill and
Wootters strategy for two-qubit case, one can extend
the concept of concurrence to a {\em concurrence
vector} defined by \begin{equation}
{\bf C}=\{\langle\psi\mid(E_\alpha -E_{-\alpha})\otimes(E_\beta
-E_{-\beta})\ket{\psi^*} \mid \alpha, \beta\in\Delta^+\}, \\
 \label{eq:concurrencevector}
\end{equation}
where $\Delta^+$ denotes the set of positive roots of $A_{N-1}$
Lie algebra. As there are totaly $N(N-1)/2$ positive roots, the
concurrence vector is a $N^2(N-1)^2/4$ dimensional vector. The
criterion for the separability of a joint pure state of bipartite
system of arbitrary dimension is that the norm of the concurrence
vector is zero, otherwise the state is entangled.

\subsection{The relation to other entanglement measurements}

In above we proposed concurrence vector on the basis of mathematics
analogy.
It is worthwhile to observe the relationship between
the afore introduced concurrence vector and
other entanglement measurement.

For a pair of qubit and qutrit which can be regarded as a pair of
spin-$1/2$ and spin-1, the concurrence
vector is a three dimensional vector given by
\begin{equation}\label{a}
\textbf{C}=\{\bra{\psi}(\sigma_{+}-\sigma_{-})
  \otimes(E_{\alpha}-E_{-\alpha})\ket{\psi^*}
   \,\mid\alpha\in\Delta^{+}\}
\end{equation}
where
$\sigma_{\pm}=(\sigma_{x}\pm i\sigma_{y})/2$, and
$\Delta^{+}$ for $A_{2}$
contains three positive roots. Thus the concurrence vector here is of three dimension.
As we known, any state of  bipartite system can be expanded as
\begin{equation}
|\psi\rangle=\sum_{\mu,j}a_{\mu j}|\mu\rangle\otimes|j\rangle,
\end{equation}
where $a_{\mu j}$ is complex coefficients, and in our present case,
$\mu=1, 2$ and $j=1, 2, 3$.
It is easy to obtain the norm of concurrence,
$
|\mathbf{C}|^2=C^2_{1}+C^2_{2}+C^2_{3}=
4(a_{11}a_{22}-a_{12}a_{21})^2
 +4(a_{12}a_{23}-a_{13}a_{22})^2+4(a_{11}a_{23}-a_{13}a_{21})^2
$.

In order to show the reliability of concurrence vector, we consider
the von Neumann entropy.
The reduced density matrix
$\rho_{A}$ and $\rho_{B}$ can be easily obtained,
\begin{eqnarray*}
\rho_{A}=aa^{\dagger}=\left(\begin{array}{ccc}
  a_{11} & a_{12} & a_{13} \\
  a_{21} & a_{22} & a_{23} \\
\end{array}
\right)
\left(
\begin{array}{cc}
  a_{11}^{\ast} & a_{21}^{\ast} \\
  a_{12}^{\ast} & a_{22}^{\ast} \\
  a_{13}^{\ast} & a_{23}^{\ast} \\
\end{array}
\right).
\end{eqnarray*}
It is a $2\times2$ matrix, thus there are two eigenvalues
$\kappa_{1}^{2}$ and $\kappa_{2}^{2}$, that are squares of the coefficients
of Schmidt decomposition
$|\psi\rangle=\kappa_{1}|x_{1}\rangle_A |y_{1}\rangle_B
 +\kappa_{2}|x_{2}\rangle_A |y_{2}\rangle_B$.
Here the
$\kappa_{1}^{2}$ and
$\kappa_{2}^{2}$ are the
roots of the following secular equation
\begin{equation}
\lambda^2-\lambda+|\mathbf{C}|^{2}/4=0,
\label{aa}
\end{equation}
where $|\mathbf{C}|$ is precisely the norm of
concurrence vector we proposed.

From Eq.(\ref{aa}), we obtain
\begin{equation}\label{solution}
\kappa_{1,2}^2=\frac{1\pm\sqrt{1-|\mathbf{C}|^2}}{2}.
\end{equation}
So the von Neumann entropy is given by
\begin{equation}
E_N(|\psi\rangle)=h((1-\sqrt{1-|\mathbf{C}|^2})/2),
\label{eq:entropy}
\end{equation}
where
\[ h(x)=-x \log_{2}x-(1-x)\log_2 (1-x).\]

On the other hand, one obtain $\rho_{B}$ by tracing out the degree
of freedom of part A, i.e.,
\begin{equation}\label{roub}
\rho_{B}=a^{\dagger}a.
\end{equation}
This is a $3\times3$ matrix whose eigenvalues are denoted by $\tilde{\kappa}_{1}^2$,
$\tilde{\kappa}_{2}^2$, $\tilde{\kappa}_{3}^2$ are roots of the algebraic equation
\begin{equation}
\lambda^3-\lambda^2+\frac{|\mathbf{C}|^2}{4}\lambda-\det(\rho_{B})=0.
\end{equation}
The reduced density matrix $\rho_{B}$ is of rank 2, i.e., $\det\rho_{B}=0$,
then there are only two non-zero eigenvalues.
The von Neumann entropy takes the same form as in
Eq.(\ref{eq:entropy}). Just like the case of
Wootters\cite{Wootters}, the von Newmann entropy here is also a
monotonous function of the norm of concurrence
vectors: $|{\bf C}|^2$.
Therefore, the concurrence vector is a reliable measurement of entanglement
of the states of qubit-qutrit system.

For bipartite qutrit system, the  general state can be written as
\begin{equation}
|\psi\rangle=\sum_{i,j}g_{ij}|i\rangle\otimes|j\rangle
\end{equation}
where $i$, $j=1, 2, 3$. The
reduced density matrix
$\rho_{B}={\rm tr}_A
|\psi\rangle \langle\psi|
\equiv g^{\dagger}g $, is
clearly a positive-definite
matrix, in which
$g=\mathrm{mat}(g_{ij})$.
Let $\kappa_1^2$,
$\kappa_2^2$,
$\kappa_{3}^2$ be the
eigenvalues of $\rho_{B}$,
which solve the following
algebraic equation
\begin{equation}
\lambda^3-\lambda^2+\frac{|\mathbf{C}|^2}{4}\lambda-\det\rho_{B}=0,
\end{equation}
where $|\mathbf{C}|^2$  is
just the square of the norm
of the concurrence vector
we proposed, namely,
\[
|\mathbf{C}|^2
=4(\kappa_1^2\kappa_2^2+\kappa_2^2\kappa_3^2+\kappa_3^2\kappa_1^2).
\]

\begin{figure}
\setlength{\unitlength}{1.0mm}
\includegraphics[width=5.2cm]{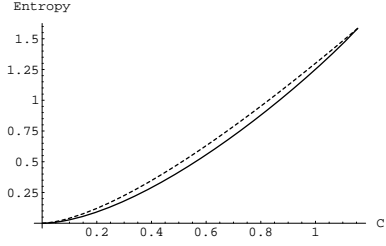}% Here is how to import EPS art
\caption{\label{fig:entropy}
The von Neumann entropy versus the norm of concurrence vector. Dash line is the supremum and
solid line the infimum}
\end{figure}

The calculation of the von
Neumann entropy exhibits an
explicit relation to the
norm of concurrence vector,
\begin{equation}
E_N(|\psi\rangle)=h\bigl(x^+, x^-\bigr), \nonumber\\
\end{equation}
where
\begin{eqnarray}
h(x^+,x^-)=-x^+\log
x^{+}-x^-\log x^- \nonumber\\
      -(1-x^+-x^-)\log (1-x^+-x^-),
\end{eqnarray}
and
\begin{eqnarray*}
x^{\pm}&=&\frac{1}{3}
 +e^{\pm i\frac{2\pi}{3}}\sqrt[3]{-\frac{q}{2}+\sqrt{\frac{q^2}{4}+\frac{p^3}{27}}}
     \\
 &+&e^{\mp i\frac{2\pi}{3}}\sqrt[3]{-\frac{q}{2}-\sqrt{\frac{q^2}{4}
  +\frac{p^3}{27}}},
      \\
p&=&\frac{|\mathbf{C}|^2}{4}-\frac{1}{3}\\
q&=&\det\rho_{B}+\frac{2}{27}-\frac{|\mathbf{C}|^2}{12}.
\end{eqnarray*}
Clearly, the von Neumann entropy depends on the norm of the
concurrence vector monotonically. Unlike the
von Neumann entropy for qubit which depends only on
the concurrence \cite{HillWootters}, it also depends on the
determinate of the reduced density matrix. The supremum (dash line) and
infimum for the von Neumann entropy versus $|C|$ are plotted in Fig.\ref{fig:entropy},
where both points of the entanglement maximum and non-entanglement coincide.
It is clearly a convex function.
%include{fig1}

The linear entropy, also a measurement of entanglement, is given by
\begin{eqnarray*}
E_L(|\psi\rangle)&=&1-tr\rho_{B}^2 \\
               &=&(\kappa_1^2+\kappa_2^2+\kappa_3^2)^2-\bigl(\kappa_1^4+\kappa_2^4+\kappa_3^4\bigr)\\
&=&2(\kappa_1^2\kappa_2^2+\kappa_2^2\kappa_3^2+\kappa_3^2\kappa_1^2)=|\mathbf{C}|^2/2
\end{eqnarray*}
which indicates a direct
relation to the norm of
concurrence vector. The
magnitude of the linear
entropy arranges from 0 to
$1-1/d$ (here $d=3$). It is
clearly a monotonically
increasing function versus
the norm of concurrence
vector. Hence the
concurrence vector we
proposed is a reasonable
measurement of entanglement for qutrits

\subsection{Qutrit via SU(3) states}

Now we consider the qutrit as a concrete example. The
states $\ket{1}$, $\ket{2}$ and $\ket{3}$ and the
corresponding weights (denoted by black dots)
$[1/2,0]$, $[-1/2,\,1/2]$ and $[0,\,-1/2]$ are
plotted in the following
\[
\setlength{\unitlength}{1mm}
\begin{picture}(33,16)(-16,-7)
 \linethickness{0.2pt}
  \put(-12,0){\dashbox{0.9}(24,0){\,}}
  \put(0,-12){\dashbox{0.9}(0,24){\,}}
  \put(0.1,0){\vector(2,3){6}} \put(7.8,7){${\bf\alpha}_2$}
  \put(0.1,0){\vector(2,-3){6}} \put(7.8,-9){${\bf\alpha}_1$}
  \put(5.6,-3.3){\circle*{1}} \put(7.2,-4.3){$\ket{1}$}
  \put(0,6){\circle*{1}}    \put(-5,5.8){$\ket{2}$}
  \put(-5.6,-3.3){\circle*{1}}\put(-11.2,-4.3){$\ket{3}$}
\end{picture}
\]
This represents the fundamental representation of SU(3). It is
well known that the direct product representation is reduced to
two irreducible representations, i.e., ${\bf 3 \otimes 3 =
3^*\oplus 6}$. The bases for the ${\bf 3^*}$ dimensional
representation are all entangled states, $\ket{\psi^-_1}$,
$\ket{\psi^-_2}$, $\ket{\psi^-_3}$, but they are not the
maximally entangled states. Additionally, it is not possible to
produce maximally entangled state by superposition in this
subspace.

Among the bases for the ${\bf 6}$ dimensional representation ( hexad),
$\ket{\psi^+_1}$,
$\ket{\psi^+_2}$ and
$\ket{\psi^+_3}$ are
entangled (not maximally
entangled) states, but the
other three $\ket{11}$,
$\ket{22}$ and
$\ket{33}$ are not
entangled states. Whereas,
three maximally entangled
states,
$\ket{\varphi_1}$,
$\ket{\varphi_2}$ and
$\ket{\varphi_3}$ can be
obtained by superposition
of those three states. All
the mentioned states are
given in the following
\begin{eqnarray}
&&\ket{\varphi_1}
 =\frac{1}{\sqrt{3}}\bigl(\ket{1 1} + \ket{2 2} + \ket{3 3}\bigr),
    \nonumber\\
&&\ket{\varphi_2}
  =\frac{1}{\sqrt{3}}\bigl(\ket{1 1}+e^{i2\pi/3}\ket{2 2}+e^{-i2\pi/3}\ket{3 3}\bigr),
    \nonumber\\
&&\ket{\varphi_3}
  =\frac{1}{\sqrt{3}}\bigl(\ket{1 1}+e^{-i2\pi/3}\ket{2 2}+e^{i2\pi/3}\ket{3 3}\bigr),
   \nonumber\\
&&\ket{\psi^\pm_1}
 =\frac{1}{\sqrt{2}}\bigl(\ket{1 2}\pm\ket{2 1}\bigr),
   \nonumber\\
&&\ket{\psi^\pm_2}
 =\frac{1}{\sqrt{2}}\bigl(\ket{2 3}\pm\ket{3 2}\bigr),
   \nonumber\\
&&\ket{\psi^\pm_3}
 =\frac{1}{\sqrt{2}}\bigl(\ket{1 3}\pm\ket{3 1}\bigr).
 \label{eq:su3 }
\end{eqnarray}
The above nine states are orthonormal to each other.
Their concurrence vectors are calculated to be
\begin{eqnarray}
\mathbf{C}_{\ket{\varphi_1}}&=&(\frac{2}{3},\, 0,\, 0,\, 0,\,
 \frac{2}{3},\, 0,\, 0,\, 0,\, \frac{2}{3}) \nonumber\\
\mathbf{C}_{\ket{\varphi_2}}&=&(\frac{2}{3}e^{-i2\pi/3},0, 0, 0,
 \frac{2}{3}, 0, 0, 0, \frac{2}{3}e^{i2\pi/3})\nonumber\\
 \mathbf{C}_{\ket{\varphi_3}}&=&(\frac{2}{3}e^{i2\pi/3}, 0, 0, 0,
 \frac{2}{3}, 0, 0, 0, \frac{2}{3}e^{-i2\pi/3})\nonumber\\
\mathbf{C}_{\ket{\psi^\pm_1}}&=&(\mp 1,\, 0,\, 0,\, 0,\, 0,\, 0,\, 0,\, 0,\, 0) \nonumber\\
\mathbf{C}_{\ket{\psi^\pm_2}}&=&(0,\, 0,\, 0,\, 0,\,  \mp 1,\, 0,\, 0,\, 0,\, 0) \nonumber\\
\mathbf{C}_{\ket{\psi^\pm_3}}&=&(0,\, 0,\, 0,\, 0,\, 0,\, 0,\, 0,\, 0,\, \mp 1)
\label{eq:concurrence-su3}
\end{eqnarray}
In addition to
$\ket{\varphi_1},\,\cdots$,
$\ket{\varphi_3}$, there
are six more orthonormal
bases,
\begin{eqnarray}
&&\ket{\varphi_4}=(\ket{12}+\ket{23}+\ket{31})/\sqrt{3},\nonumber\\[1mm]
&&\ket{\varphi_5}=(\ket{12}+e^{i2\pi/3}\ket{23}+e^{-i2\pi/3}\ket{31})/\sqrt{3},\nonumber\\
&& \;\;   \cdots, \nonumber\\
&&\ket{\varphi_9}=(\ket{21}+e^{-i2\pi/3}\ket{32}+e^{i2\pi/3}\ket{13})/\sqrt{3},\nonumber
\end{eqnarray}
which were adopted as a generalization of EPR pairs
when discussing teleportation \cite{Bennet93}. The
concurrence vectors for those states are easily
calculated, for example,
$\mathbf{C}=(0, 2/3, 0, 0, 0, -2/3, -2/3, 0, 0)$ for
$\ket{\varphi_4}$.
Actually, all the states $\ket{\varphi_l}$,
($l=4,\cdots,\,9$) are shown to be maximally
entangled by calculating the norm of their
concurrence vectors $|{\bf C}|^2=4/3$.

\subsection{Qutrit via SO(3) states}

The fundamental representation of SU(2) is
carried out by spin-1/2 system which refers to the
qubit. We know the Bell bases,
$\ket{\psi^B_\pm}=\bigl(\ket{\uparrow\downarrow}
\pm\ket{\downarrow\uparrow}\bigr)/\sqrt{2}$,
$\ket{\varphi^B_\pm}=\bigl(\ket{\uparrow\uparrow}
\pm\ket{\downarrow\downarrow}\bigr)/\sqrt{2}
$ are simultaneous
eigenstates of $S_x\otimes
S_x$ and $S_z\otimes S_z$,
which is not valid for
spin-1 (i.e., SO(3)
representation) or high
spin systems. This is
because $S_x\otimes S_x$,
$S_y\otimes S_y$ and
$S_z\otimes S_z$ commute to
each other only for
spin-1/2 case. We consider
such a pair of qutrits that
their direct product
representation decomposes
into SO(3) irreducible
representations, i.e.,
${\bf 3 \otimes 3 = 5
\oplus 3 \oplus 1}$. Some
of their states read,
\begin{eqnarray}
&&\ket{\chi^\pm_1}
 =\frac{1}{\sqrt{2}}\bigl(\ket{0 1}\pm\ket{1 0}\bigr),
   \nonumber\\
&&\ket{\chi^\pm_{-\!1}}
 =\frac{1}{\sqrt{2}}\bigl(\ket{-\!1 0}\pm\ket{0-\!\!1}\bigr),
   \nonumber\\
&&\ket{\chi^\pm_0}
  =\frac{1}{\sqrt{4\pm 2}}\bigl(\ket{-\!1 1}\pm \ket{0 0}
  +\ket{0 0}\pm\ket{1-\!\!1}\bigr),
   \nonumber\\
&&\ket{\chi^0_0}
  =\frac{1}{\sqrt{3}}\bigl(\ket{-\!1 1}-\ket{0 0}+\ket{1-\!\!1}\bigr),
   \nonumber\\
&&\ket{\phi^\pm}
 =\frac{1}{\sqrt{2}}\bigl(\ket{1 1}\pm\ket{-\!1-\!\!1}\bigr).
 \label{eq:so3 }
\end{eqnarray}
It is easy to calculate their concurrence vectors:
\begin{eqnarray}\label{eq:vecso3}
\mathbf{C}_{\ket{\chi^\pm_1}}&=&(\mp 1,\, 0,\, 0,\, 0,\, 0,\,0,\, 0,\, 0,\, 0)\nonumber \\
\mathbf{C}_{\ket{\chi^\pm_{-1}}}&=&(0,\, 0,\, 0,\, 0,\, \mp 1,\, 0,\, 0,\, 0,\, 0)\nonumber \\
\mathbf{C}_{\ket{\chi^-_0}}&=&(0,\, 0,\, 0,\, 0,\, 0,\, 0,\, 0,\, 0,\, 1)\nonumber \\
\mathbf{C}_{\ket{\chi^+_0}}&=&(0,-\frac{2}{3}, 0,-\frac{2}{3}, 0, 0, 0, 0,-\frac{1}{3})\nonumber \\
\mathbf{C}_{\ket{\chi^0_0}}&=&(0,\frac{2}{3}, 0,\frac{2}{3}, 0, 0, 0, 0,-\frac{2}{3})\nonumber \\
\mathbf{C}_{\ket{\phi^\pm}}&=&(0,\, 0,\, 0,\, 0,\, 0,\,0,\, 0,\, 0,\pm1)
\end{eqnarray}
The evaluation of the norm
of concurrence vector
indicates that the singlet
$\ket{\chi^0_0}$ is
maximally entangled $|{\bf
C}|^2=4/3$; the triplet,
$\ket{\chi^-_j}$
(hereafter for SO(3)
$j=-1,0,1$) are entangled
but not maximally entangled
$|{\bf C}|^2=1$; among the
pentads only
$\ket{\chi^+_j}$ are
entangled states while
$\ket{11}$ and
$\ket{-1-1}$ are
unentangled states whose
superposition provides the
two entangled states,
$\ket{\phi^\pm}$.

\section{Concurrence vector for mixed states}\label{sec:mix}

In the light of our extension of the measurement of entanglement in
terms of concurrence vector for pure states,
it is natural to introduce
\begin{equation}\label{eq:tilde}
\ket{\tilde{\psi}}_{\alpha \beta}=(E_{\alpha}-E_{-\alpha})
\otimes(E_{\beta}-E_{-\beta})\ket{\psi^*},
\end{equation}
here $\alpha$ and $\beta$ refer to the aforementioned positive roots of
$A_{N-1}$ Lie algebra.
Similar to the strategy of Wootters\cite{Wootters},
we define some matrices given by
\begin{equation}
\tau^{\alpha\beta}_{ij}=\bra{v_i} (E_{\alpha}-E_{-\alpha})
\otimes(E_{\beta}-E_{-\beta})\ket{v_j^*}
\end{equation}
where $\{\ket{ v_i} \mid i=1,\,2,\cdots \}$ are the eigenvectors
of the density matrix $\rho$ which characterizes a given mixed state.
Apparently, the matrices $(\tau_{ij})$ are symmetric.
According to Takagi's factorization \cite{horn}, for any symmetric
matrix $A$, there exists a unitary $U$ and a real nonnegative diagonal matrix
$d=\mathrm{diag}(\lambda_1,\ldots,\lambda_n)$ such that $A=U d~ U^{T}$ and the diagonal
entries of $d$ are the nonnegative square roots of the corresponding eigenvalues of
$AA^{\ast}$.
Thus there exists a decomposition
$\ket{x_i}=U_{ij}^*\ket{v_j}$ satisfying
\begin{eqnarray}
\bra{x_i}(E_\alpha-E_{-\alpha})\otimes(E_\beta-E_{-\beta})\ket{x^*_j}\nonumber\\
=(U\tau^{\alpha\beta} U^{T})_{ij}=\lambda_{i}^{\alpha\beta}\delta_{ij}.
\end{eqnarray}
Here $\lambda^{\alpha\beta}_i$ is the nonnegative square root of the eigenvalue of
$\tau^{\alpha\beta} \tau^{\ast \alpha\beta}$ and
\begin{eqnarray}
\tau^{\alpha\beta}\tau^{\alpha\beta\ast}=
\sqrt{\rho}(E_{\alpha}-E_{-\alpha})\otimes \hspace{10mm}\nonumber\\
(E_{\beta}-E_{-\beta})\rho^{\ast}
(E_{\alpha}-E_{-\alpha})\otimes(E_{\beta}-E_{-\beta})\sqrt{\rho}.
\end{eqnarray}

One can make another decomposition of $\rho$ in terms of $\{\ket{y_i}\}$,
\begin{equation}
|y_1\rangle=|x_1\rangle;\ \ |y_j\rangle=i|x_j\rangle,\quad j\neq 1
\end{equation}
If one repeat the steps given by Wootters in Ref.\cite{Wootters}, one can show that
for some given positive roots $\alpha$, $\beta$,
the concurrence is expressed as
$$C^{\alpha\beta}=\max\{ 0,\, \lambda_1^{\alpha\beta}-\sum_{i=2}^{n}\lambda_i^{\alpha\beta} \}$$
where $\lambda^{\alpha\beta}_1=\max\{\lambda^{\alpha\beta}_i,i=1,\ldots, N\}$.
Thus we derive a formula to calculate concurrence vector for mixed
state. Then the norm of such a concurrence vector can be employed to measure the
entanglement of mixed state:
\begin{equation} \label{eq:mixed-concurrence}
|\mathbf{C}|^2=\sum_{\alpha\beta}|C_{\alpha\beta}|^2
\end{equation}

For SU(2) case there is only one positive root, which corresponds to a pair of qubits,
the concurrence vector (\ref{eq:mixed-concurrence}) is one dimensional, which is just
the original definition of Wootters' concurrence.
The concurrence vector in Eq. (\ref{eq:mixed-concurrence}) is expected to study the pairwise
entanglement of SU(3) Heisenberg model similar to the strategy of XXZ chain \cite{Gu03} which
is in progress.

\section{Concurrence surface and entanglement edge}\label{sec:surface}

Calculating the norm of the concurrence vectors for an
arbitrary normalized state in the three-dimensional subspace
spanned by either
$\{\ket{\psi^-_1}, \,\ket{\psi^-_2},\,\ket{\psi^-_3}
\}$, or
$\{\ket{\psi^+_1}, \,\ket{\psi^+_2},\,\ket{\psi^+_3}
\}$, or
$\{\ket{\varphi_1}, \,\ket{\varphi_2},\, \ket{\varphi_3} \}$
respectively,  we obtain
several conclusions. Any
states in the space of the
former two cases yield a
constant norm of
concurrence vectors, which
means that the Hilbert
subspace manifests a fixed
entanglement. The norm of
concurrence vector vanishes
in the third case when the
expanding coefficients take
some particular magnitudes.
This indicates that some
states in that Hilbert
subspace are separable.
Braunstein {\it et al.}\
\cite{Braunstein99}
analyzed the separability
of $N$-qubit states near
the maximally mixed state.
Vidal and Tarrach
\cite{Vidal99} give a
separability boundary for
the mixture of the
maximally mixed state with
a pure state. Caves and
Milburn \cite{Caves00}
discussed the lower and
upper bounds on the size of
the neighborhood of
separable states around the
maximally mixed state of
qutrits. One will see in
the following that all
these become easily
understandable by making
use of the concept of
concurrence vectors.

The entanglement features of those three Hilbert subspaces discussed
previously arise from the sign properties of the
concurrence vectors.
Considering a Hilbert subspace constituted by some bases obeying
\begin{equation}
\inn{\psi_\mu}{\tilde{ \psi_\nu}}_{\alpha\beta} \propto \delta_{\mu\nu}
\end{equation}
where $\ket{\tilde{ \psi_\nu}}_{\alpha\beta} $ is defined by Eq.(\ref{eq:tilde}).
One immediate criterion is that the state vectors lie in the Hilbert subspace are
all entangled as long as the corresponding nonzero components of
the concurrence vectors for those bases have the same signs
({\it i.e.}, positive or negative), which implies impossible to make
a superposition state with all the components of the concurrence vector
vanish.

\begin{figure}
\setlength{\unitlength}{1.5mm}
\includegraphics[width=3.9cm]{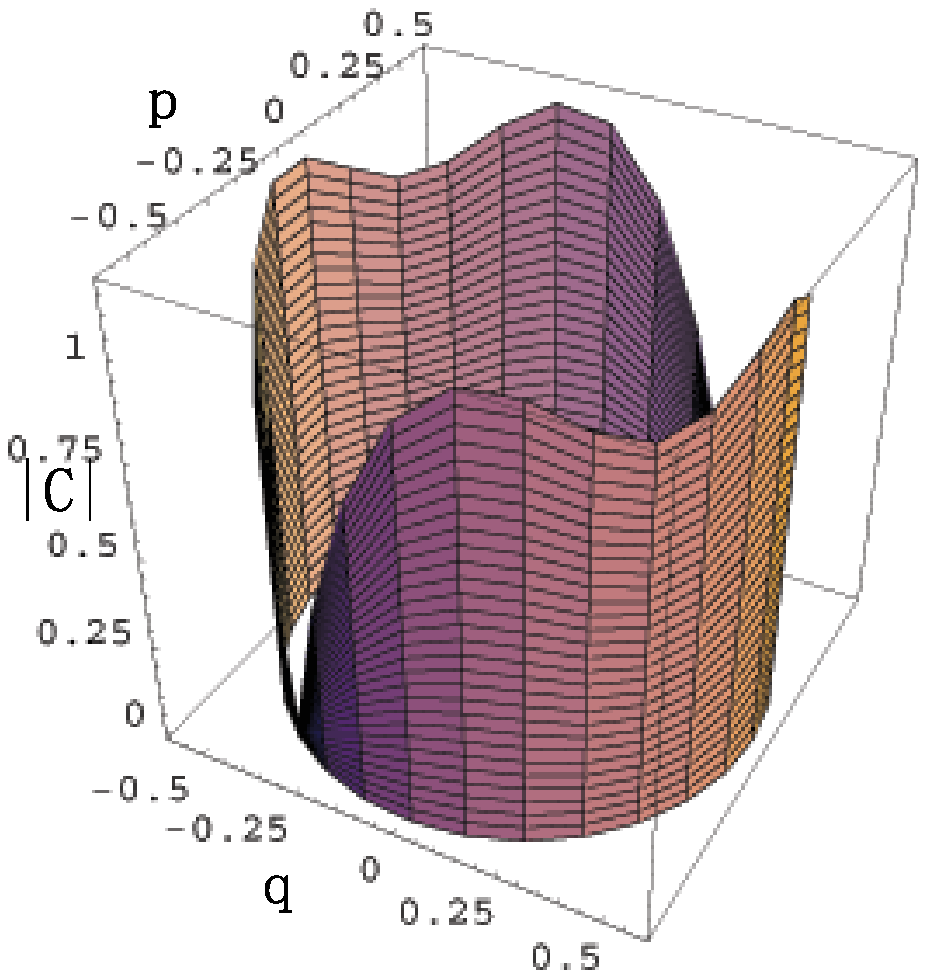}% Here is how to import EPS art
\includegraphics[width=4.2cm]{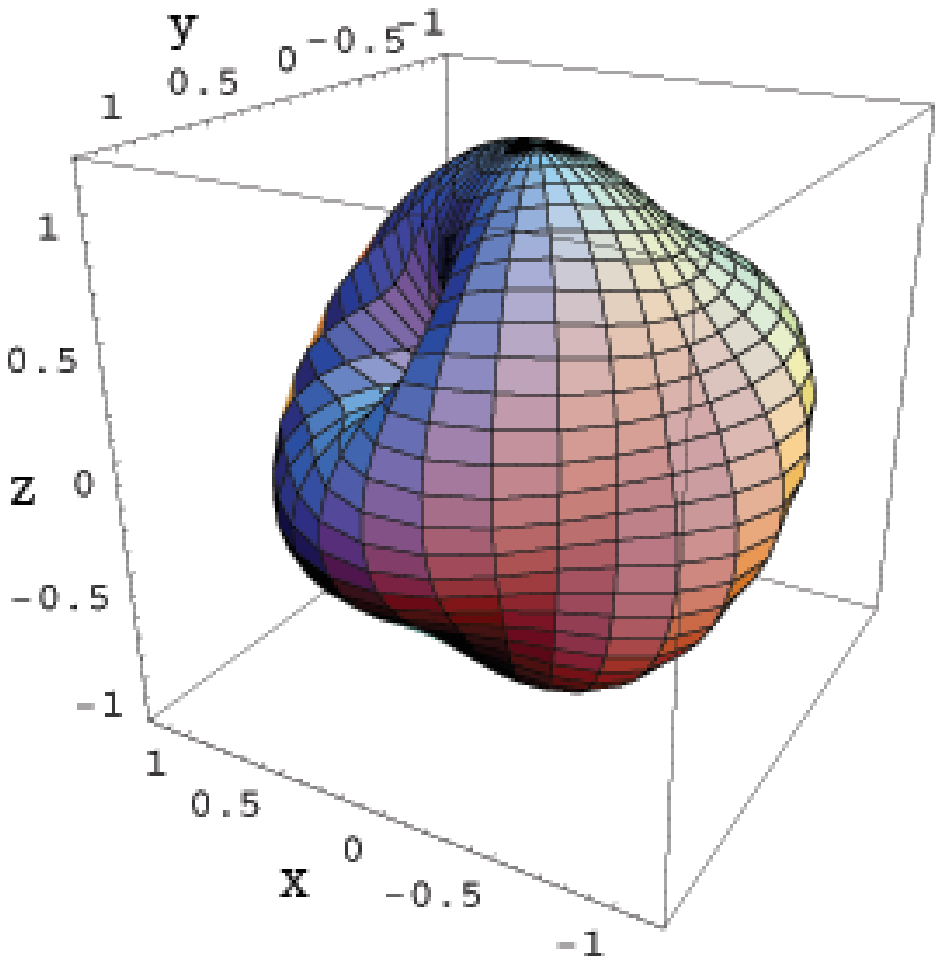}
\caption{\label{fig:su3}(Color online) The left is the relation of $\mid{\bf
C}\mid$ versus $p$ and $q$ for state $p(\ket{\psi^+_1}+
\ket{\psi^+_2}+\ket{\psi^+_3})
  +q(\ket{11}+ \ket{22} + \ket{33})$.
The right is the surface $|{\bf C}(\theta,\,\phi)|$ for Hilbert
subspace $\sin\theta\cos\phi\ket{\varphi_1}
 + \sin\theta\sin\phi\ket{\varphi_2}
 +\cos\theta\ket{\varphi_3})$}
\end{figure}

It is instructive to observe the entanglement edges in
the subspace of the $6$ dimensional representation
of SU(3). Since the norm of concurrence vectors are
constant in the subspace spanned by
$\ket{\psi^+_j}$, we let the coefficients of these
three bases equal to $p$. We also choose the
coefficients for
$\ket{11}$,
$\ket{22}$ and
$\ket{33}$ to be the
same $q$ so that to plot a three dimensional picture.
The curve of the norm of concurrence vector versus
$p$ and $q$ is given in Fig.(\ref{fig:su3}), the
left. When
$p\rightarrow\sqrt{2}q$ and
$q\rightarrow 1/\sqrt{3}$
it approaches the entanglement edge.

Now we consider the case
for SO(3). As nonzero
components of concurrence
vectors for the triplet are
positive and that for  the
three entangled bases in
the pentad are negative,
both the whole Hilbert
subspace spanned by
$\{\ket{\chi^-_j}\}$ and
the subspace spanned by
$\{\ket{\chi^+_j}\}$ are
entangled. Because the
nonzero components of
concurrence vectors for
$\ket{\phi^+}$ is
positive but that for
$\ket{\phi^-}$ is
negative (see Eq.(\ref{eq:vecso3})), the five
dimensional Hilbert
subspace spanned by
$\{\ket{\chi^+_j}\}$ and
$\ket{\phi^\pm}$ are not
fully entangled.

\begin{figure}
\setlength{\unitlength}{1.5mm}
\includegraphics[width=4.1cm]{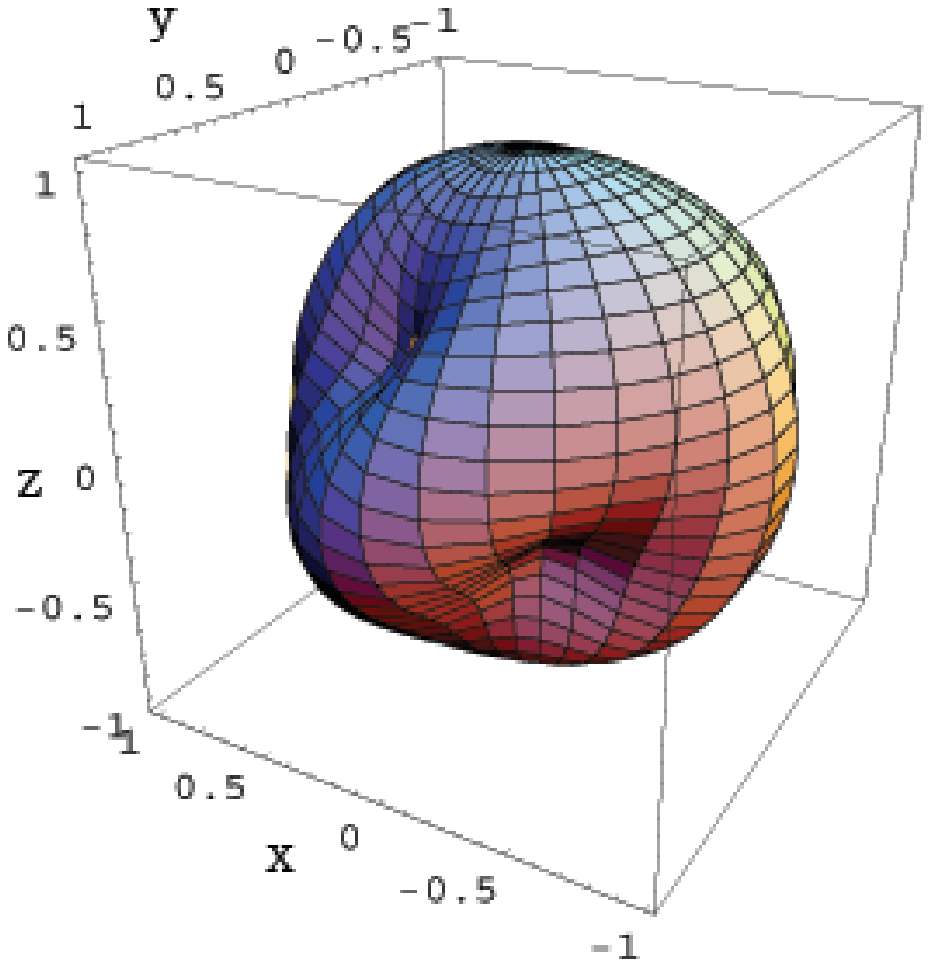}% Here is how to import EPS art
\includegraphics[width=4.2cm]{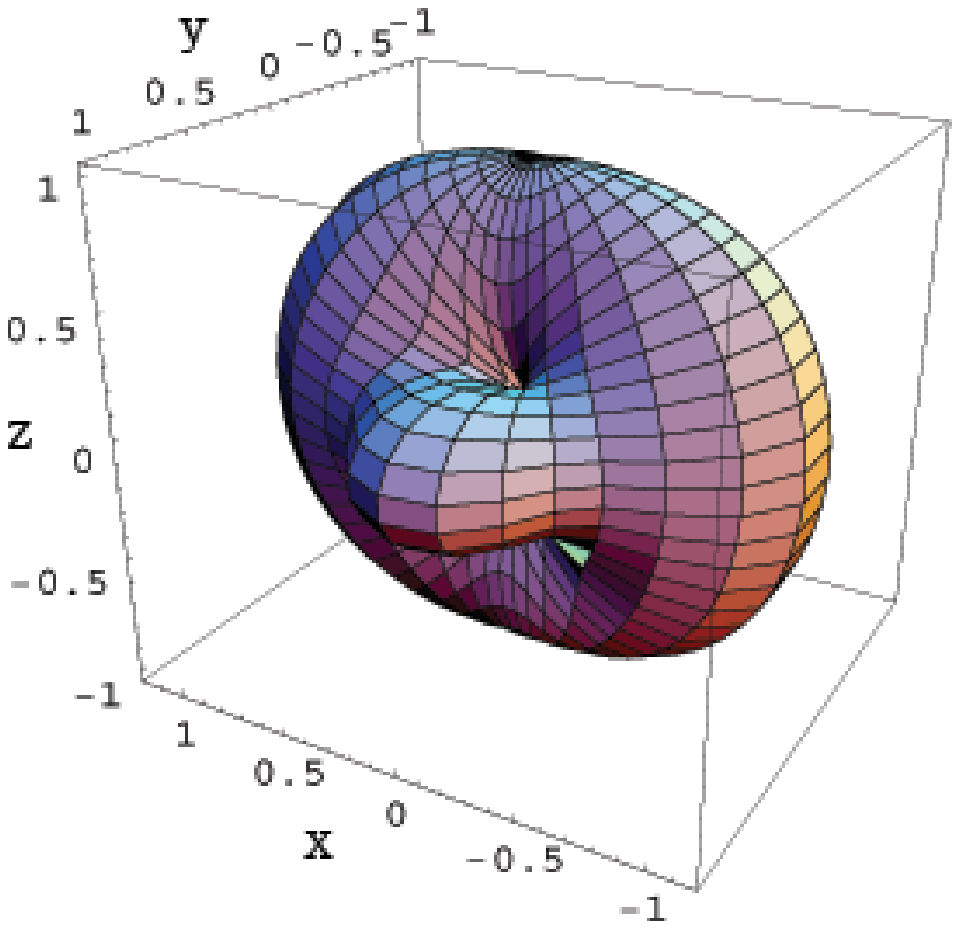}
\caption{\label{fig:so3} (Color online)
The surfaces $\mid{\bf
C}(\theta,\,\phi)\mid$ for
the Hilbert subspaces
$\sin\theta\cos\phi\ket{\chi^+_1}
 + \sin\theta\sin\phi \ket{\chi^+_0}
 +\cos\theta\ket{\chi^+_{-1}}$ (left),
  and
  $\sin\theta\cos\phi\ket{\chi^0_0}
 + \sin\theta\sin\phi \ket{\phi^+}
 +\cos\theta\ket{\phi^-}$ (right).
}
\end{figure}

The parameter space of
general states in some
three dimensional Hilbert
subspace is described by a
two-sphere
$(\theta,\,\phi)$ and a
phase factor $e^{i\delta}$
which is fixed to unit
without loss of generality.
It is illuminative to
observe the geometry
structures by plotting the
norm of concurrence vector
as radial coordinate, i.e.,
$r(\theta,\,\phi)=\mid{\bf
C}_{\ket{\theta,\,\phi}}\mid$,
we call it {\em concurrence
surface}. The concurrence
surfaces for the triplets
of either SO(3) or SU(3) is
simply a sphere of unit
radius, which encloses a
spheroid of volume
$4\pi/3$. The concurrence
surface for the Hilbert
subspace spanned by the
three maximally entangled
states $\{\varphi_j\}$ of
SU(3) is plotted in
Fig.\ref{fig:su3} (the
right), whose enclosure
merely occupies $13.5\%$
more volume than a unit
spheroid does though their
bases are maximally
entangled. This is
obviously due to the
presence of entanglement
edges. The concurrence
surfaces for the Hilbert
subspace spanned by the
three originally entangled
states in the pentad (left
figure), and that spanned
by the singlet together
with $\ket{\varphi^\pm}$ are plotted in Fig.\ref{fig:so3} (right
figure). Their enclosures occupy volumes of $3.18868$
and $2.75916$ respectively,
the later is smaller though
one of its bases is
maximally entangled.

%include{fig2}

%include{fig3}

Quantum entanglement implies correlations
between the results of measurements on component
subsystems of a larger physical system, which can
not be understood by means of correlations between
local classical properties inherent in those
subsystems. Wang and Zanardi \cite{WangZanardi02}
studied the entanglement of unitary operators on a
bipartite quantum systems that is related to the
entangling power of the associated quantum
evolutions. It maybe useful to associate those topics
with the evolution of concurrence vectors.
Zanardi and Rasetti \cite{Zandardi99} showed
that the notion of generalized Berry phase can
be used for enabling quantum computation, which
also supports the necessity of the investigation on the
parameter manifold related to quantum system with high
rank symmetries.

\section{Summary}

In above we proposed a concurrence vector to
measure the entanglement of bipartite system of
arbitrary dimension by employing the fundamental
representation of $A_n$ ($n=N-1$) Lie algebra.
For pure state, we have discussed the relation between the concurrence vector
and the von Neumann entropy for both qubit-qutrit pair and
qutrit-qutrit pair. We also gave a formula to calculate
the norm of concurrence vector for mixed states on the basis of the
strategy of Wootters.
We have shown that the norm of concurrence vector can be
used to evaluate the entanglement of a state,
i.e., a separable state has a vanishing norm.
However, we have not yet found an example such that the entanglement of UPB
state or PPT state with no product vectors in its range can be detected by
calculating concurrence vector.
Another advantage of concurrence vectors is easy to judge
whether a Hilbert subspace spanned by some entangled bases is fully entangled
states, or there exists entanglement edge in the subspace. If the
corresponding nonzero components of concurrence vectors of the basis states
are all positive or all negative, all the state
vectors lie in this Hilbert subspace are entangled.
We calculated the concurrence vectors for the states
related to SU(3) and SO(3) explicitly. We also
discussed concurrence surface and compared the
volumes enclosed by the surface for various cases.
Their geometry will be useful for understanding
the entanglement capacities in various Hilbert
subspaces.

\section*{Acknowledgement}

The work is supported by NSFC Grant No.10225419.

\appendix*
\section{}

As the qutrit can be related by the generators of
the SU(3) group and the $N$-dimensional Hilbert
space by that of the SU(N) group.
According to Cartan-Weyl analysis, the generators can be divided
into two sets: the Cartan subalgebra which is the maximal Abelian
subalgebra, and the remaining generators which play the similar
role as the above $J_\pm$. The structure constants for the
commutation relations of those operators can be described by the
called root space diagram. The root space diagram for $A_2$ Lie
algebra has a hexagonal shape,
\[
\setlength{\unitlength}{1mm}
\begin{picture}(38,20)(-15,-10)
 \linethickness{0.5pt}
  \put(0,0){\circle{2}}\put(0,0){\circle{1}}
  \put(0.2,0){\vector(1,0){12}}   \put(13.5,-0.3){${\bf\alpha}_1+{\bf\alpha}_2$}
  \put(-0.2,0){\vector(-1,0){12}} \put(-28.5,-0.3){$-{\bf\alpha}_1-{\bf\alpha}_2$}
  \put(0.1,0){\vector(2,3){6}}    \put(6.5,10){${\bf\alpha}_2$}
  \put(-0.1,0){\vector(-2,-3){6}} \put(-8.6,-12){$-{\bf\alpha}_2$}
  \put(-0.1,0){\vector(-2,3){6}}  \put(-8.8,10){$-{\bf\alpha}_1$}
  \put(0.1,0){\vector(2,-3){6}}   \put(6.5,-12){${\bf\alpha}_1$}
\end{picture}
\]
The double circle at the center indicates the existence of two
generators $H_1$, $H_2$ in the Cartan subalgebra. Unlike the spin
systems whose states are labelled by the eigenvalue of $J_z$ which
is the called magnetic quantum number, the states of SU(3) system
are labelled by the eigenvalues of $(H_1,\,H_2)$ that is therefore
a vector called weight vector.   We adopt nonorthogonal bases to
expend the root vectors choosing the {\em simple roots}
${\bf\alpha}_1$ and ${\bf\alpha}_2$ as coordinate bases, which is
clear and convenient for physicists. Placing contravariant
components, $({\bf\alpha})^i$ in conventional parenthesis, we have
the positive roots ${\bf\alpha}_1=(1,\,0)$,
${\bf\alpha}_2=(0,\,1)$, ${\bf\alpha}_1 + {\bf\alpha}_2=(1,\,1)$,
and the negative roots $-{\bf\alpha}_1=(-1,\,0)$,
$-{\bf\alpha}_2=(0,\,-1)$, $-{\bf\alpha}_1 -
{\bf\alpha}_2=(-1,\,-1)$. If placing covariant components,
$({\bf\alpha})_j$, in square parenthesis, we easily obtain from
the above root space diagram that ${\bf\alpha}_1=[ 1,\,-1/2\,]$,
${\bf\alpha}_2=[ -1/2,\,1\,]$, ${\bf\alpha}_1 + {\bf\alpha}_2=[
1/2,\,1/2\,]$, $-{\bf\alpha}_1=[ -1,\,1/2\,]$, $-{\bf\alpha}_2=[
1/2,\,-1\,]$, and $-{\bf\alpha}_1 - {\bf\alpha}_2=[ -1/2,\,-1\,]$.
Then one can easily write out the following commutation relations
\begin{eqnarray}
&\bigl[H_i ,\, H_j\bigr]=0,\hspace{10mm} & \,\nonumber\\
&\bigl[H_j,\,E_{\bf\alpha}\bigr]=({\bf\alpha})_j E_{\bf\alpha},
  &\bigl[E_{\bf\alpha},\,E_{-\bf\alpha}\bigr]=2({\bf\alpha})^i H_i,
   \nonumber\\
&\bigl[E_{\bf\alpha},\, E_{\bf\beta}\bigr]=
E_{\bf\alpha+\beta}\hspace{4mm}
  & \;{\rm if} \;\; {\bf\alpha+\beta}\in\Delta.
 \label{eq:commutators}
\end{eqnarray}
where $\Delta$ denotes the
set of nonzero roots of
nonexceptional Lie algebra.
The above commutation
relations imply that
$E_{\pm\alpha_j}$ play the
roles of raising/lowering
operators like $J_\pm$ of
the angular momentum
operator. Moreover, there
are more than one
operators, $H_j$'s, that
commute to each other. For
SU(N) case which
corresponds to $A_{N-1}$
Lie algebra, there are
$N-1$ generators in the
Cartan subalgebra, hence
$i,j=1,..., N-1$ and
Eq.(\ref{eq:commutators})
also fulfil. Due to  $N-1$
dimensional vectors are
difficult to depict, the
root space diagram is
represented graphically by
a two dimensional diagram,
called Dynkin diagram:
\[
\setlength{\unitlength}{1mm}
\begin{picture}(38,5)(2,-2)
\linethickness{0.5pt}
\put(2,2){\circle{2}}\put(3,2){\line(1,0){7.8}}
 \put(1,-2){$\alpha_1$}
\put(12,2){\circle{2}}\put(13,2){\line(1,0){7.8}}
 \put(11,-2){$\alpha_2$}
\put(22,2){\circle{2}}\put(23,2){\line(1,0){2.6}}
 \put(21,-2){$\alpha_3$}
       \put(26.6,1){$\cdots$}\put(31.4,2){\line(1,0){2.5}}
\put(35,2){\circle{2}}\put(36,2){\line(1,0){7.8}}
  \put(34,-2){$\alpha_{N-2}$}
\put(45,2){\circle{2}}
 \put(45,-2){$\alpha_{N-1}$}
\end{picture}
\]
where each open dot "$\circ$" denotes a simple root, the angle
between a pair of simple roots is $120^o$ if a line connecting
them;  that is $90^o$ if no line connecting them. Their covariant
components of simple roots are easily calculated from the Dynkin
diagram,
\begin{eqnarray}
&&\alpha_1 = [1,\, -1/2,\;\; 0,\hspace{3mm}\cdots\hspace{6mm},0], \nonumber\\
 &&\alpha_2 =[-1/2,\,1,\,-1/2,\, 0,\cdots,0], \nonumber\\
  && \;\;\cdots, \nonumber\\
  &&\alpha_{N-1}= [0,\,0, \cdots, 0,\,-1/2,\,1 \,].
\end{eqnarray}
The simple roots are normalized to unity so that
the structure constants in Eq. (\ref{eq:commutators})
differs from the Cartan matrix in textbook of group
theory by a factor $1/2$. The advantage of our
convention is that if returning to the SU(2),
whose root space diagram has a simple line shape,
\[
\setlength{\unitlength}{1mm}
\begin{picture}(28,0)(-8,0)
 \linethickness{0.5pt}
  \put(0,0){\circle*{1.5}}
  \put(0.2,0){\vector(1,0){12}}   \put(13,-2){${\bf\alpha}$}
  \put(-0.2,0){\vector(-1,0){12}} \put(-16.8,-2){$-{\bf\alpha}$}
\end{picture}
\]
the eigenvalues of $S_z$ are $1/2$ and $-1/2$ respectively for
spin up and down, otherwise, they would be $1$ and $-1$.

We consider an $N$-dimensional Hilbert space which carries out the
fundamental representation of Lie algebra $A_{N-1}$, the whole
weight vectors that label the state bases can be easily produced
from the highest weight vector $[1/2, 0,\,\cdots]$ by Weyl
reflection which is easily realized by means of the covariant
component of simple roots, i.e., \\[2mm]
$ [\frac{1}{2},\,0,\,\cdots ]\shortstack{ -$\alpha_1$ \\
 $\longrightarrow$ }
[-\frac{1}{2},\,\frac{1}{2},0,\,\cdots ]\shortstack{$-\alpha_2$ \\
$\longrightarrow $}\cdots
\shortstack{$-\alpha_{N-1}$\\
$\longrightarrow$} [0,\cdots, -\frac{1}{2}].
$ \\[1mm]
\noindent
 Consequently, the $N$ state vectors are generated by the
lowering operators.
\begin{eqnarray}
\ket{1}\shortstack{$E_{-\alpha_1}$ \\ $\longrightarrow$}
\ket{2}\shortstack{$E_{-\alpha_2}$ \\ $\longrightarrow$}
\cdots\ket{N-1} \shortstack{$E_{-\alpha_{N-1}}$ \\
$\longrightarrow$} \ket{N}.
 \label{eq:state}
\end{eqnarray}
The positive simple roots, whereas, just give the reverse of the
above relations.

\end{document}